\documentclass[article]{elsarticle}

\journal{Journal of \LaTeX\ Templates}

\usepackage{comment}
\usepackage{amsmath}
\usepackage{amssymb}
\usepackage{hyperref}
\usepackage{cleveref}
\usepackage{comment}
\usepackage{caption}

\newcommand{\SPE}{SPE}
\newcommand{\MPE}{MPE}

\begin{document}

\begin{frontmatter}

\title{Investigation of magnetic-field effects on photomultiplier tubes for the first Large-Sized Telescope of the Cherenkov Telescope Array Observatory} 

\cortext[correspondingauthor]{Corresponding author}

\author[icrr]{Yukiho Kobayashi\corref{correspondingauthor}}
\author[saitama]{Dai Tateishi}
\author[saitama]{Yuji Sunada}
\author[saitama]{Takuto Kiyomoto}
\author[saitama]{Nobuaki Sasaki}

\author[tokai]{Kazuki Abe}
\author[cppm]{Franca Cassol}
\author[yamagata]{Shuichi Gunji}
\author[icrr]{Daniela Hadasch}
\author[cppm]{Dirk Hoffmann}
\author[cppm]{Julien Houles}
\author[icrr]{Yusuke Inome}
\author[ibaraki]{Hideaki Katagiri}
\author[icrr]{Hidetoshi Kubo}
\author[tokai]{Junko Kushida}
\author[icrr,mpp]{Daniel Mazin}
\author[konan]{Masaya Mizote}
\author[saitama]{Tsutomu Nagayoshi}
\author[yamagata]{Takeshi Nakamori}
\author[tokai]{Kyoshi Nishijima}
\author[icrr]{Seiya Nozaki}
\author[icrr]{Hideyuki Ohoka}
\author[isee,kmi]{Akira Okumura}
\author[tokushima]{Reiko Orito}
\author[icrr]{Takayuki Saito}
\author[icrr]{Shunsuke Sakurai}
\author[isee]{Mitsunari Takahashi}
\author[saitama]{Yukikatsu Terada}
\author[icrr,mpp]{Masahiro Teshima}
\author[konan]{Tokonatsu Yamamoto}
\author[ibaraki]{Tatsuo Yoshida}

\address[icrr]{Institute for Cosmic Ray Research, University of Tokyo, 5-1-5, Kashiwa-no-ha, Kashiwa, Chiba 277-8582, Japan}
\address[saitama]{Graduate School of Science and Engineering, Saitama University, 255 Simo-Ohkubo, Sakura-ku, Saitama city, Saitama 338-8570, Japan}
\address[tokai]{Department of Physics, Tokai University, 4-1-1, Kita-Kaname, Hiratsuka, Kanagawa 259-1292, Japan}
\address[cppm]{Aix Marseille Univ, CNRS/IN2P3, CPPM, Marseille, France}
\address[yamagata]{Department of Physics, Yamagata University, 1-4-12 Kojirakawa-machi, Yamagata-shi 990-8560, Japan}
\address[ibaraki]{Faculty of Science, Ibaraki University, 2 Chome-1-1 Bunkyo, Mito, Ibaraki 310-0056, Japan}
\address[mpp]{Max-Planck-Institut für Physik, Boltzmannstraße 8, 85748 Garching bei München, Germany}
\address[konan]{Department of Physics, Konan University, 8-9-1 Okamoto, Higashinada-ku Kobe 658-8501, Japan}
\address[isee]{Institute for Space-Earth Environmental Research, Nagoya University, Chikusa-ku, Nagoya 464-8601, Japan}
\address[kmi]{Kobayashi-Maskawa Institute (KMI) for the Origin of Particles and the Universe, Nagoya University, Chikusa-ku, Nagoya 464-8602, Japan}
\address[tokushima]{Graduate School of Technology, Industrial and Social Sciences, Tokushima University, 2-1 Minamijosanjima, Tokushima 770-8506, Japan}

\begin{abstract}
Photomultiplier tubes (PMTs) are widely used in imaging atmospheric Cherenk-ov telescopes. Their response can depend on the telescope pointing direction through changes in their orientation relative to the geomagnetic field. Such a dependence was indicated during the calibration campaign of the CTAO LST-1. In this work, the LST-1 calibration data are analyzed to quantify the gain dependence in terms of the geomagnetic field, and dedicated laboratory measurements are performed to test the hypothesis that the observed dependence originates from magnetic-field effects. Both the on-site and laboratory measurements show a consistent dependence of the PMT gain on the magnetic field, while the laboratory measurements further separate the effects on the gain and the excess noise factor. These results provide a clearer understanding of the gain variation observed in LST-1 in terms of magnetic-field effects.
\end{abstract}

\begin{keyword}
Photomultiplier tubes; Magnetic-field effects; Calibration; Cherenkov telescopes
\end{keyword}

\end{frontmatter}

\section{Introduction}
Photomultiplier tubes (PMTs) are widely used in imaging atmospheric Cher-enkov telescopes (IACTs) such as H.E.S.S.~\cite{HESS}, MAGIC~\cite{MAGIC}, VERITAS~\cite{VERITAS}, and Cherenkov Telescope Array Observatory (CTAO)~\cite{LST_Module}.
It is well known that the PMT response is affected by magnetic fields, even at geomagnetic-field strengths, due to the deflection of photoelectrons between the photocathode and the first dynode~\cite{CALVO2010, LEONORA2013}.
In IACTs, the orientation of PMTs relative to the geomagnetic field changes with the telescope pointing direction, unlike in many ground-based and underground experiments, where PMTs are fixed with respect to the ground.
Consequently, the geomagnetic-field effects on the PMT response should be evaluated over the range of telescope pointing directions encountered during operation, which can be done directly using on-site data.
During the calibration campaign of the first Large-Sized Telescope (LST-1) of CTAO, a dependence of the PMT response on the telescope pointing direction was observed, suggesting an effect of the geomagnetic field on the PMT response~\cite{LST1CamCal}.
While the magnitude of the observed variation is negligible for the calibration of LST-1, the possible magnetic-field origin of the variation was not examined in detail in the previous study.

In this work, we further investigate the effect observed in the LST-1 data in terms of the geomagnetic field and quantify its impact.
In addition, we perform dedicated laboratory measurements to achieve a deeper understanding of the PMT response to magnetic fields.
These measurements also allow us to separately determine the PMT gain and the excess noise factor, which cannot be disentangled using the on-site data alone.
\section{On-site Test}

The effects of the geomagnetic field on the LST-1 PMTs were evaluated on site during the camera calibration campaign.
This section describes the LST-1 camera and the PMT gain calibration method, and presents the results of dedicated tests assessing the geomagnetic field effects.

\subsection{LST-1 Camera}
The focal plane camera of LST-1 consists of 1855 PMTs, which are arranged in a hexagonal geometry.
The PMTs (R11920-100-20), developed in collaboration with Hamamatsu Photonics K.K., were specially designed for the CTAO-LST project and are employed in the LST-1 camera~\cite{LSTPMT2016, LSTPMT2017}.
The R11920-100-20 PMTs feature a fast response with a pulse width of $\sim3$\,ns~\cite{LST_Module}.
Seven PMTs are mounted to a single readout board, and this unit is treated as one PMT module.
The light concentrators are installed at the PMT entrance windows to enhance the light collection efficiency~\cite{Lightguide2017}.
Signals from the PMTs are amplified through the PreAmplifier for the CTA cameras (PACTA) and sent to the readout board~\cite{sanuy2012}.
The readout board is equipped with Domino Ring Sampler version 4 (DRS4) chips~\cite{Ritt2008}, in which the PMT signals are sampled at a frequency of 1 GHz and recorded as waveforms.
The sampled waveforms are then digitized, read out from the modules, and stored on the disk.

\begin{figure}
    \centering
    \includegraphics[width=0.9\textwidth]{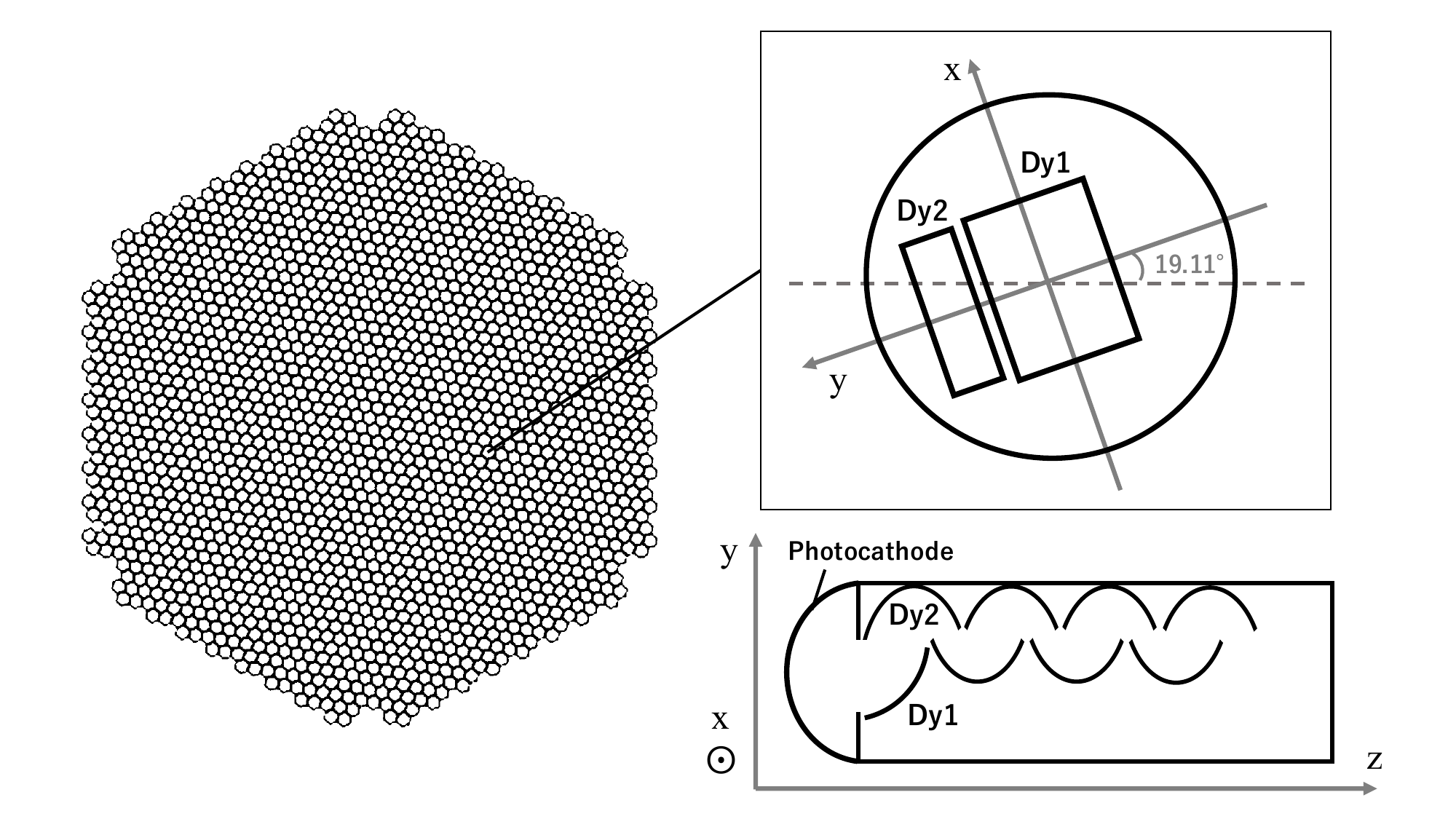}
    \caption{The front view of the LST-1 camera (left) and a sketch of the PMT geometry in the camera, as seen from the photocathode perspective (top right) and from the side of the PMT (bottom right). The labels Dy1 and Dy2 denote the first and second dynodes, respectively.}
    \label{fig:PMT-geometry}
\end{figure}

Figure~\ref{fig:PMT-geometry} shows the hexagonal arrangement of the LST-1 PMTs in the camera and their internal geometry, including the first and second dynodes.
All PMTs are installed with the same rotational orientation about their axis, thereby aligning the first and second dynodes in the same direction across the camera.
A local $x$–$y$ coordinate system is defined for each PMT in the plane perpendicular to the PMT tube axis.
As shown in \Cref{fig:PMT-geometry}, the $y$-axis is aligned with the direction connecting the first and second dynodes, and the $x$-axis is orthogonal to it.
The $y$-axis is tilted by 19.11$^\circ$ from the horizontal axis of the entire camera.
Each PMT is surrounded by the magnetic shield made of mu-metal to minimize the effects of the geomagnetic field on the PMT response~\cite{LST_PMT_Inome}.

\subsection{PMT Gain Calibration Method}

The gain calibration of the LST-1 PMTs is performed using the so-called excess-noise-factor method~\cite{f-factor-method}.
This method estimates the PMT gain based on the charge distribution obtained from a stable light source.
When the means and variances of the signal (pedestal) charge distribution are denoted by $\overline{Q} (\overline{\rm ped})$ and $\sigma^2_Q (\sigma^2_{\rm ped})$, respectively,
the PMT gain can be estimated as 
\begin{equation}
G_F = \frac{1}{F^2}\left(\frac{\sigma^2_Q - \sigma^2_{\rm ped}}{\overline{Q} - \overline{\rm ped}} - B^2 (\overline{Q} - \overline{\rm ped})\right),
\label{eq:F-factor-method}
\end{equation}
where $B$ represents the systematic variation in the signal charge and typically has a value of $\sim$0.03 for the LST-1 camera, while $F$ is the so-called excess noise factor~\cite{LST1CamCal}.
The excess noise factor is defined as
\begin{equation}
F^2 = 1 + \left(\frac{\sigma_G}{G}\right)^2,
\label{eq:f-factor}
\end{equation}
where $G$ is the average PMT gain and $\sigma_G$ is the standard deviation of the PMT gain.
It characterizes the statistical fluctuations in the PMT amplification and is an intrinsic property of the PMT.
For the LST-1 PMTs, an average value of $F^2=1.22$ was measured and is adopted for the calibration of all PMTs~\cite{LST1CamCal}.

A calibration flasher, referred to as CaliBox, is installed at the center of the telescope mirror dish~\cite{CaliBox}.
CaliBox is designed to produce a uniform flash over the camera.
The flash has a pulse width of $\sim$1\,ns, as measured in the laboratory, and a wavelength of 355\,nm.
By examining the distributions of the extracted charge from a number of flasher pulses, the gain of each PMT can be estimated following \Cref{eq:F-factor-method}.
This method has notable advantages that the calibration can be performed with a large amount of light and that the calibration results can be monitored even during observations by interleaving flasher data with regular air-shower data.

\subsection{Measurement of the Geomagnetic-field Effect}

To evaluate the effects of the geomagnetic field on the calibration results, dedicated calibration data were taken at different telescope pointing directions so that the magnetic-field components in the PMT coordinate system varied.
The data were recorded for all combinations of zenith angles of 5$^\circ$, 35$^\circ$, and 75$^\circ$ and azimuth angles of 0$^\circ$, 90$^\circ$, 180$^\circ$, and 270$^\circ$.
Data were also obtained during telescope slewing to reduce data gaps between different pointing directions.
Both flasher and pedestal data were recorded periodically at a rate of 1\,kHz during data taking.

The excess-noise-factor method was applied to the calibration data, and the PMT gain was estimated for each accumulated data set containing more than $10^4$ flasher pulses and pedestal samples.
The data analysis was performed using \texttt{lstcam\_calib}~\cite{cassol_2025_15729582}, a Python package developed for the LST camera calibration, with the standard configurations described in~\cite{LST1CamCal}.

The strength and the direction of the geomagnetic field at the telescope site were obtained from the 14th Generation of the International Geomagnetic Reference Field model~\cite{IGRF14}.
The site coordinates were assumed to be a latitude of 28.762$^\circ$, a longitude of $-17.892^\circ$, and an altitude of 2170\,m above sea level.
The date of interest was 11 September 2020, when the test was performed.
The resulting geomagnetic field strength was 38726\,nT, with a declination of $-5.038^\circ$ and an inclination of 37.447$^\circ$.
Using the geomagnetic field model and the PMT geometry shown in Fig.~\ref{fig:PMT-geometry}, the geomagnetic field was transformed into the local PMT coordinate system for each telescope pointing direction. The dependence of the calibration results on the derived magnetic-field components was then examined.

\subsection{Results and Discussion} \label{sec:onsite_results}

\begin{figure}
    \centering
    \includegraphics[width=0.8\textwidth]{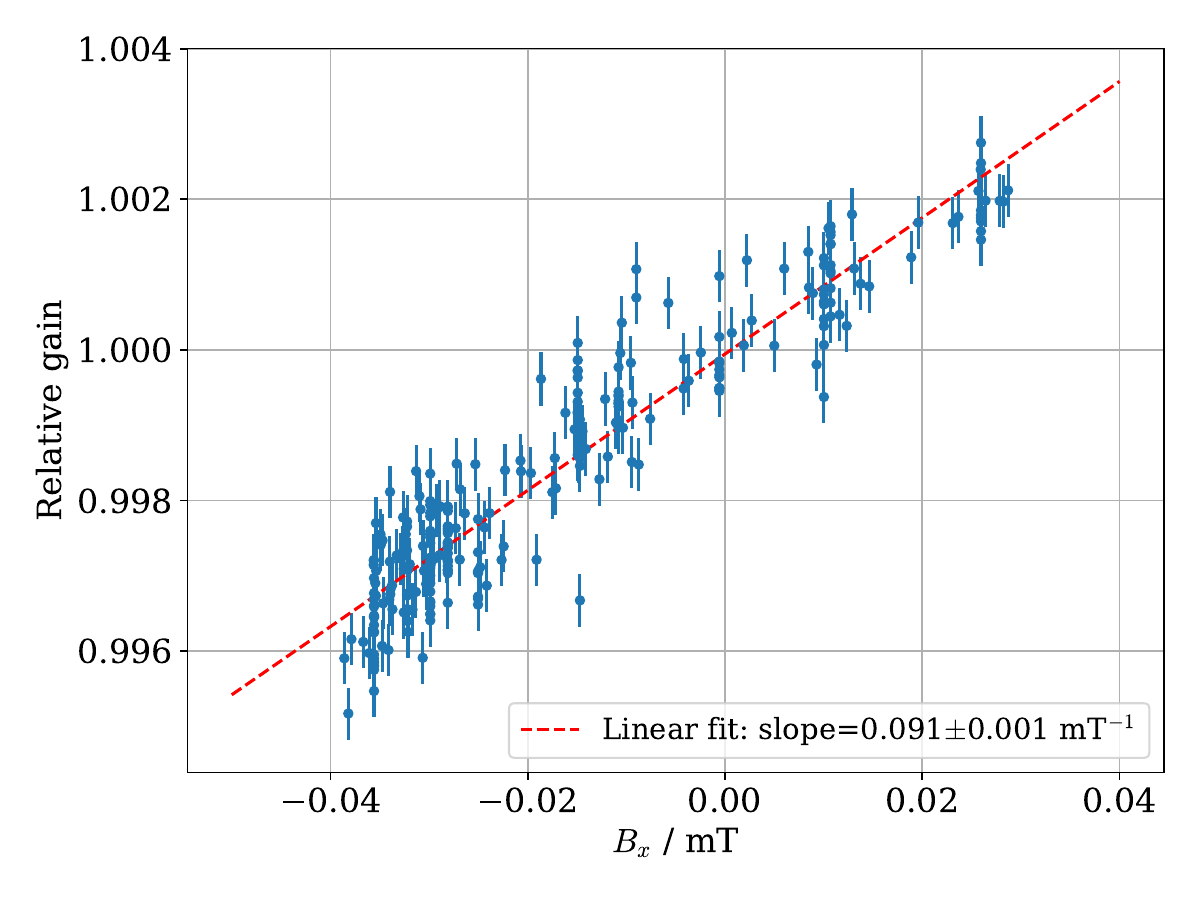}
    \caption{Dependence of the pixel-averaged PMT gain on the expected x-component of the geomagnetic field experienced by the PMTs. The average PMT gain is normalized at $B_x=0$. The red dashed line indicates a linear fit to the data. The error bars represent statistical error only.}
    \label{fig:GainvsBx_LST}
\end{figure}

\Cref{fig:GainvsBx_LST} shows the relative variation of the pixel-averaged PMT gain as a function of the x-component of the geomagnetic field experienced by the PMTs, $B_x$, expected at the time of data acquisition.
The dependence on the other magnetic-field components, $B_y$ and $B_z$, is discussed in \ref{sec:GainvsByBz}.
A clear correlation is observed between the PMT gain and the expected $B_x$.
A linear fit to the data indicates a slope of $0.091 \pm 0.001\,\mathrm{mT}^{-1}$ for the relative PMT gain.

The gain variation shown in~\Cref{fig:GainvsBx_LST} can be interpreted as an effect of the geomagnetic field on the PMT response.
The magnetic field can bend the trajectories of photoelectrons inside the PMTs, particularly those between the photocathode and the first dynode, which can lead to changes in the gain.
It is reasonable that the x-component of the geomagnetic field has a dominant effect because, when $B_x$ is present, the momentum of photoelectrons is deflected along the $y$-axis, where the dynode structure is asymmetric, as illustrated in~\Cref{fig:PMT-geometry}.

\begin{figure}
    \centering
    \includegraphics[width=0.9\textwidth]{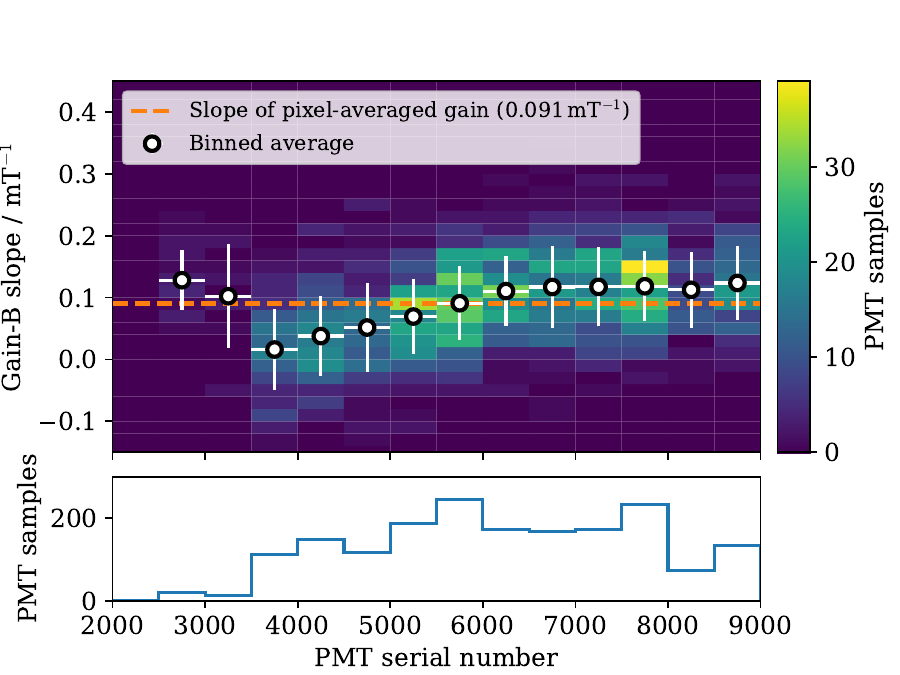}
    \caption{Top: distribution of the PMT samples as a function of the PMT serial number (numeric part) and the slope of the linear fit to the PMT gain as a function of the magnetic field. The white points represent the average slope in each serial number bin, with error bars showing the standard deviation within each bin. The orange dashed line indicates the average dependence, corresponding to the red dashed line in~\Cref{fig:GainvsBx_LST}. Bottom: distribution of PMT serial numbers in the analyzed sample.}
    \label{fig:SlopeVsSerial}
\end{figure}

The gain dependence on the magnetic field is also examined for each pixel.
The slope obtained from a linear fit to the data in each pixel is shown in~\Cref{fig:SlopeVsSerial} as a function of the PMT serial number.
The figure shows a systematic trend of the slope increasing with the serial number.
Since the serial numbers correspond to the order in which the PMTs were produced, the results may suggest changes in the manufacturing process during mass production that affect the magnetic-field dependence.
For each serial number bin, the obtained slope has a standard deviation of $\sim 0.06\,\mathrm{mT}^{-1}$.
Subtracting the typical statistical uncertainty in the fitted slope, $\sim 0.05\,\mathrm{mT}^{-1}$, this corresponds to a variation of $\sqrt{0.06^2 - 0.05^2}\,\mathrm{mT}^{-1} \sim 0.03\,\mathrm{mT}^{-1}$ among individual PMTs.
These variations are relevant when interpreting the results from the laboratory measurements with a limited number of PMT samples, as discussed in~\Cref{sec:lab_results}.
The possible dependence of the geomagnetic-field effect on the PMT position within the camera is examined in \ref{sec:SlopecVsDist}.

Given that the gain dependence on $B_x$ is estimated to be $0.091 \pm 0.001$\,mT$^{-1}$ and that the geomagnetic field strength at the site is 0.039\,mT, the maximum possible gain variation due to telescope rotation is expected to be $\sim$0.7\%.
This effect does not have a significant impact on the camera calibration, as it is much smaller than the total uncertainty budget for the camera photo-detection efficiency $\sim10$\%.
In addition, the excess-noise-factor method allows the PMT gain to be calibrated on the fly, naturally accounting for the effects of the geomagnetic field.
Therefore, the observed gain variation does not pose a problem for the calibration, provided that it reflects a genuine change in the PMT gain.

Nevertheless, dedicated laboratory measurements were required to reproduce the observed magnetic-field dependence under controlled conditions and to gain a deeper understanding of the effects of magnetic fields on the PMT response.
These measurements also allow the magnetic-field dependence of the excess noise factor to be investigated.
If the excess noise factor itself is affected by the magnetic field, it would introduce an additional systematic uncertainty in the on-site gain calibration, but such a dependence cannot be examined using the on-site data alone.
The laboratory measurements are described in the following section.
\section{Laboratory Measurement}
In order to further investigate the effects of magnetic fields on the PMT response under controlled conditions, laboratory measurements were carried out.
These measurements aim to validate whether the gain variations observed in the LST-1 data can be fully attributed to geomagnetic-field effects and to disentangle the dependence of the PMT gain and the excess noise factor on the magnetic field.

\subsection{Measurement Setup}
\begin{figure}
    \centering
    \includegraphics[width=1.0\textwidth]{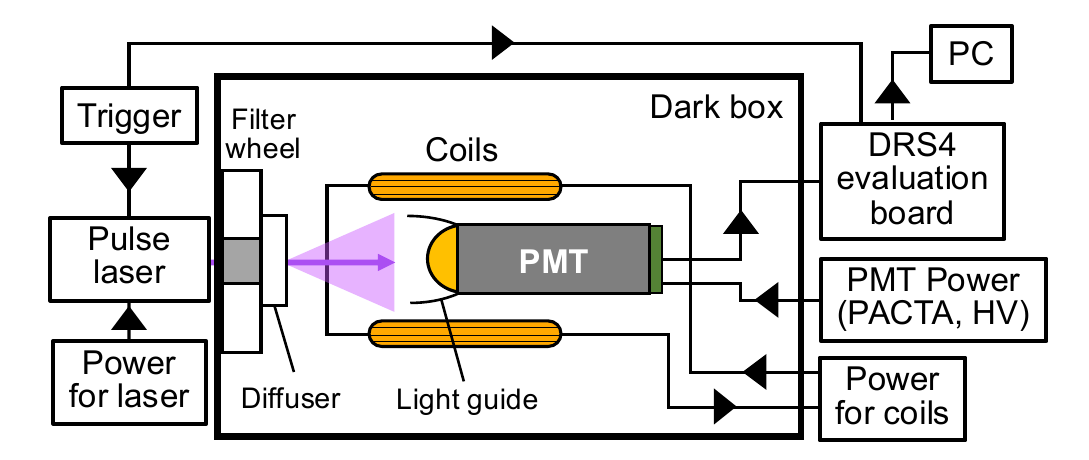}
    \includegraphics[width=0.6\linewidth]{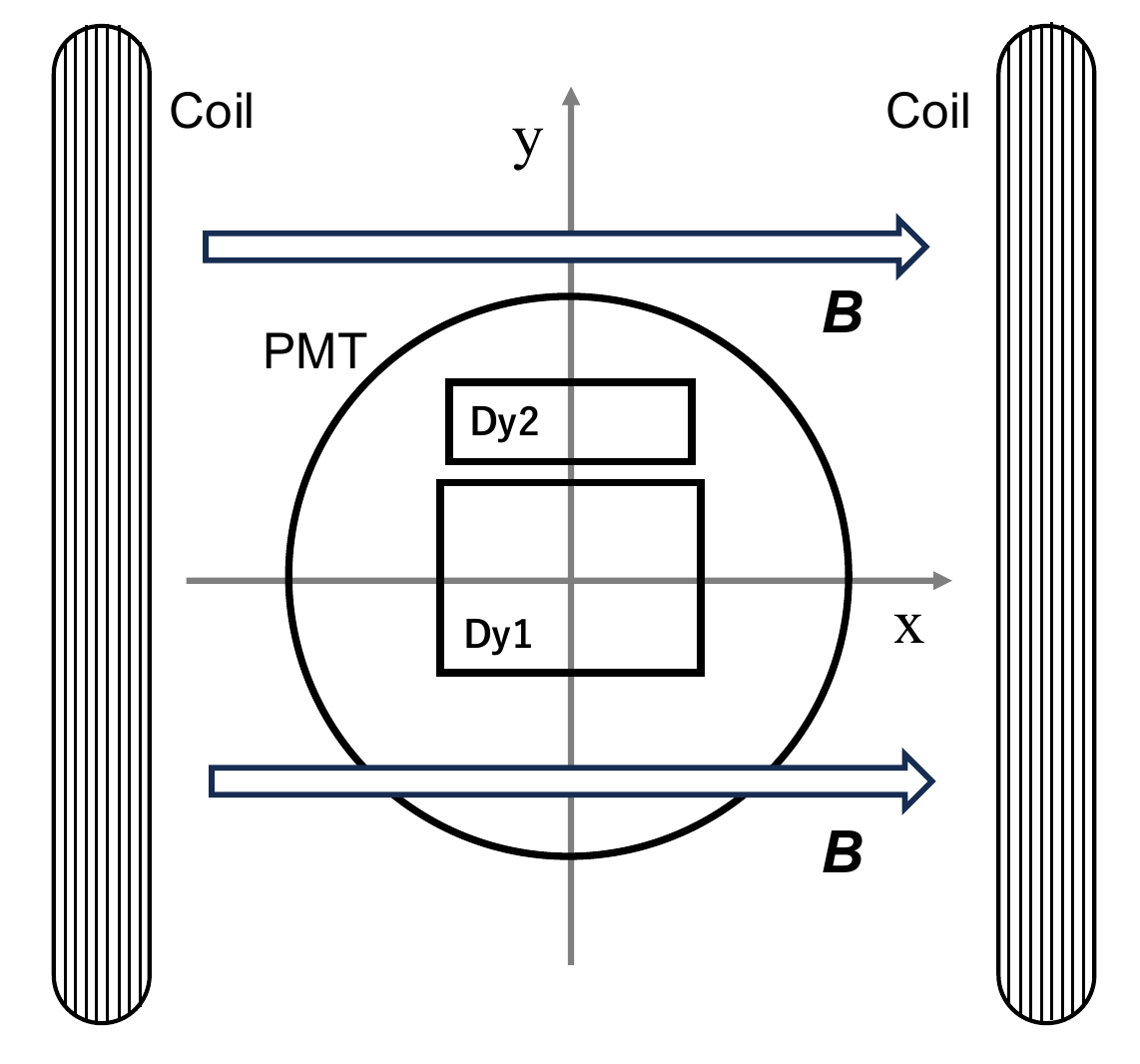}
    \caption{Top: schematic overview of the measurement setup. Bottom: front-view schematic of the PMT geometry and magnetic-field configuration.}
    \label{fig:setup}
\end{figure}

A schematic view of the measurement setup is shown in~\Cref{fig:setup}.
The measurements were performed in a dark box. The magnetic field was generated by Helmholtz coils with a radius of $r=10.5$\,cm and a winding number of $N=200$. A PMT was placed between the coils, with its photocathode positioned on the central axis of the coils, such that the magnetic field was most uniform around the photocathode and the first dynode, where the effects of the magnetic field are expected to be most significant.
The PMT was aligned such that its internal $x$-axis was parallel to the central axis of the coils, allowing the magnetic field to be applied along the x-direction, where the dominant effects were suggested by the LST-1 data.
The calibration of the magnetic field generated by the coils is described in~\ref{sec:magcal}.
The ambient geomagnetic field along the $x$-axis was measured to be $B_0 = -0.01$\,mT, and all magnetic-field strengths quoted for the laboratory measurements include this contribution.

The PMT was illuminated by a pulsed laser with a pulse width of 70\,ps~\cite{Inome2019}. The PMT signal was first amplified on the PACTA and then recorded as a waveform using a DRS4 chip, sampled at a frequency of 5\,GHz. Data acquisition with the DRS4 chips was performed using a DRS4 evaluation board~\cite{Ritt2008}.

\subsection{Data Acquisition and Analysis}
Two different types of measurements were performed: multi-p.e. (\MPE) and single-p.e. (\SPE) measurements.
The \MPE\ response was measured to examine the reproducibility of the on-site LST-1 results, while the \SPE\ response was measured to disentangle the dependence of the PMT gain and the excess noise factor on the magnetic field.
Five samples of R11920-100-20 PMTs were measured under identical conditions.
Their serial numbers are ZQ4058, ZQ5628, ZQ6295, ZQ7568, and ZQ7980.
For each measurement, 1000 dark events without laser pulses were recorded by using the thickest filter in the filter wheel to block the laser light. The resulting average waveform baseline was used to correct possible systematic baseline variations.

\subsubsection{\MPE\ Measurement}
The response of R11920-100-20 PMTs to a large amount of light was measured to investigate the dependence of the excess-noise-factor method results on the magnetic field and to compare them with the on-site LST-1 results.
For each PMT, a nominal high voltage (HV) in the range of 1030\,V to 1080\,V, depending on the PMT, was applied.
This voltage yields the nominal gain, defined as $4 \times 10^4$ for the LST-1 PMTs.
The laser intensity was set to $\sim120$ p.e. per pulse.

The magnetic field was scanned once from the maximum positive value (0.32\,mT) to the maximum negative value ($-0.34$\,mT), and then back from the lowest to the highest value.
This magnetic field range is approximately eight times larger than the geomagnetic field strength at the LST-1 site.
At each magnetic field strength, $2 \times 10^5$ events were acquired during each of the downward and upward scans.

The signal was extracted from the waveforms by summing the samples in the time window from $\langle t \rangle - 4\,\mathrm{ns}$ to $\langle t \rangle + 8\,\mathrm{ns}$, where $\langle t \rangle$ denotes the average pulse peak position over all events in each acquisition at a given magnetic-field strength.
The pedestal was evaluated from waveform samples preceding the signal, using a time window of the same width.
The PMT gain was then estimated from the signal and pedestal charge distributions using \Cref{eq:F-factor-method}.
The systematic $B$ term was omitted because the $B$ constant was not evaluated in this study. The possible effects of this approximation are discussed in \Cref{sec:lab_results}.
The excess noise factor was assumed to be $F^2=1.22$, the representative value for the LST-1 PMTs.
The results from the upward and downward scans were averaged at each magnetic field strength. 

\subsubsection{\SPE\ Measurement}
The \SPE\ response of R11920-100-20 PMTs was measured to separately  evaluate the effects of the magnetic field on the PMT gain and the excess noise factor.
An HV of 1500\,V, higher than the nominal operating voltage, was applied to all PMTs in order to achieve a better signal-to-noise ratio and to enable a detailed investigation of the \SPE\ charge distribution.
Because the voltage at the first dynode of the LST-1 PMTs is fixed to 350\,V by a Zener diode, the excess noise factor, which is dominated by the amplification at the first dynode, is expected to be largely independent of the applied high voltage~\cite{LST_Module}.
The laser intensity was adjusted to $\sim0.3$\,p.e. per pulse.
The magnetic field was scanned similarly to the \MPE\ measurement, over a field range from $-0.58$\,mT to 0.55\,mT.
At each magnetic field strength, $5 \times 10^4$ events were acquired during each of the upward and downward scans.

\begin{figure}
    \centering
    \includegraphics[width=0.8\textwidth]{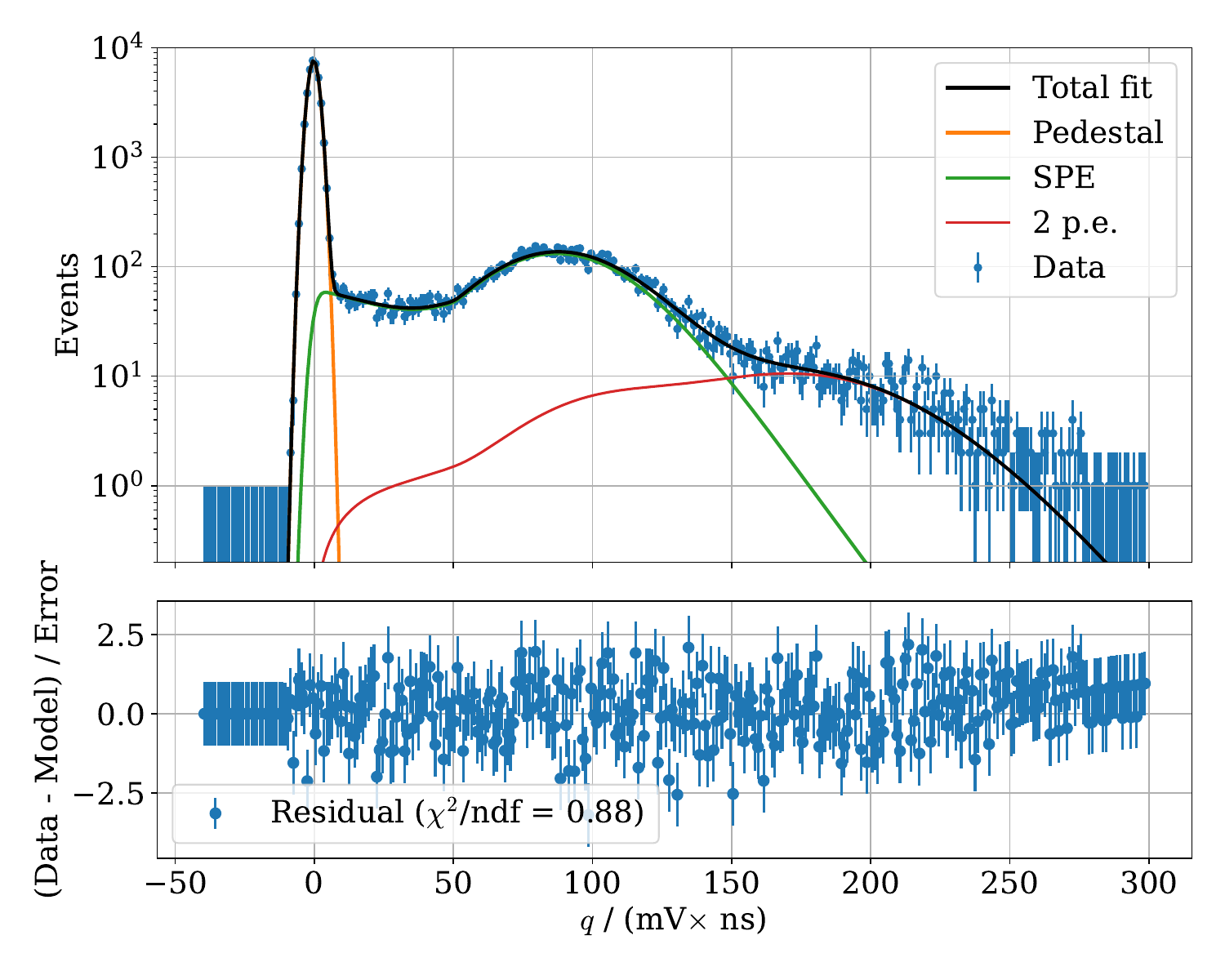}
    \caption{Example charge spectrum fitted with the model in~\Cref{eq:spe_model_all}. The PMT sample is ZQ4058, and the data were taken with no magnetic field applied by the coils. Top: charge spectrum (blue points) fitted with the model (black curve). The pedestal, SPE, and 2~p.e. components are shown by the orange, green, and red curves, respectively. Bottom: fit residuals normalized by the statistical uncertainty of each data point.}
    \label{fig:SPEFit}
\end{figure}

The signal was extracted from the waveforms using a time window of $\pm4$\,ns around the average pulse peak position over all events in each acquisition.
The resulting charge distribution was then fitted to extract the \SPE\ charge spectrum.
The fit model and procedure follow those presented in~\cite{LST_Module}.
Details are given in~\ref{sec:spe_model}.
An example charge spectrum together with the fitted model is shown in~\Cref{fig:SPEFit}.

The PMT gain $G$ is computed as the average charge weighted by the fitted SPE model $m_1(q)$.
The excess noise factor $F^2$ is calculated using \Cref{eq:f-factor}, where $\sigma_G$ is obtained as the standard deviation of the SPE model $m_1(q)$.
Since analytical calculations of the uncertainties in the best-fit values of $G$ and $F^2$ are too complicated, the uncertainties are evaluated by randomly varying the fitting parameters around their best-fit values and determining the largest deviation while maintaining the goodness of fit. 
The random sampling assumes that each parameter follows a Gaussian distribution whose width is given by the uncertainty in its best-fit value. 
The covariance matrix obtained from the fit is used to take into account correlations among the parameters.
The goodness-of-fit criterion is defined as
\begin{equation}
\chi^2 < \max\left(\chi^2_{68},\, \chi^2_{\rm best}+(\chi^2_{68}-1)\right),
\end{equation}
where $\chi^2$ denotes the reduced $\chi^2$ of the fit, $\chi^2_{68}$ is the 68th percentile of the expected $\chi^2$ distribution, and $\chi^2_{\rm best}$ is the $\chi^2$ value for the best fit.
The second condition, $\chi^2 - \chi^2_{\rm best} < \chi^2_{68} - 1$, is introduced to avoid underestimating the uncertainties by allowing deviations of up to $\chi^2_{68}-1$ from $\chi^2_{\rm best}$. This is particularly necessary when the best fit does not yield a sufficiently good $\chi^2_{\rm best}$, possibly due to model imperfections or systematic uncertainties in the data.
The resulting uncertainty estimates are therefore expected to be conservative.
Finally, the results from the upward and downward scans were averaged at each magnetic field strength.

\subsection{Results and Discussion} \label{sec:lab_results}

In this section, the results of the laboratory measurements are presented for both \MPE\ and \SPE\ measurements.

\subsubsection{\MPE\ Measurement}

\begin{figure}
    \centering
    \includegraphics[width=0.9\textwidth]{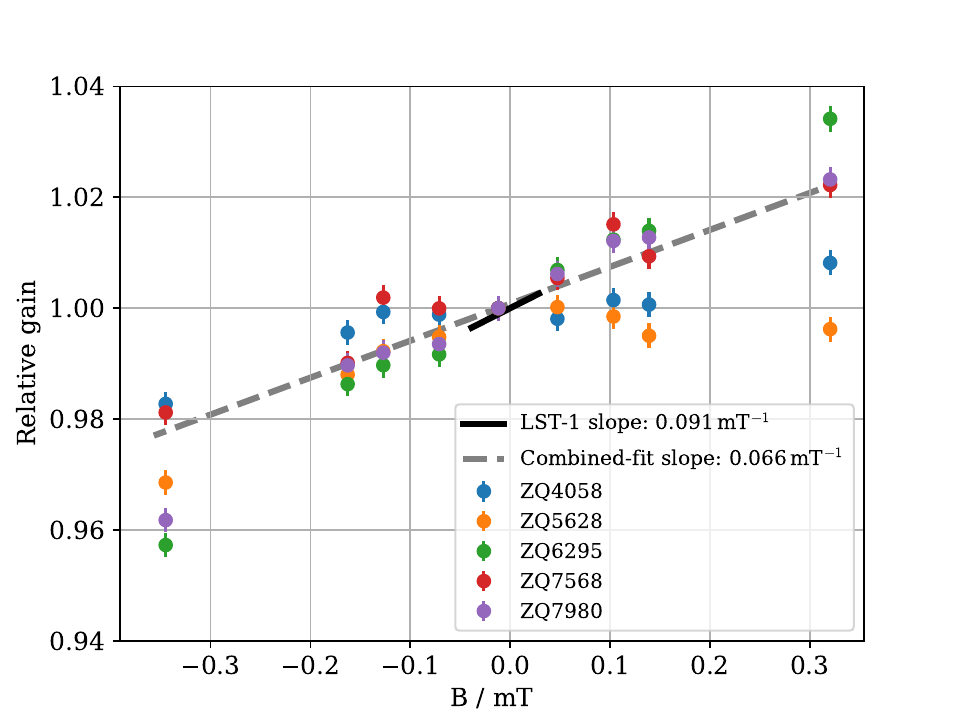}
    \caption{Dependence of the PMT gain, $G_F$, obtained from the \MPE\ measurements using the excess-noise-factor method, on the magnetic field strength. For each PMT, the gain is normalized to the value obtained with no magnetic field applied by the coils. The gray dashed line indicates a linear fit to the data using all PMT samples. The dependence observed in the LST-1 on-site data is indicated by the black line.}
    \label{fig:MPEGain_vs_B}
\end{figure}

\Cref{fig:MPEGain_vs_B} shows the dependence of the PMT gain, estimated using the excess-noise-factor method, on the magnetic field strength.
All PMT samples show a positive correlation between the gain and the magnetic field strength.
The linear-fit slope of the magnetic-field dependence for each PMT sample is summarized in the ${\rm d}G_F/{\rm d}B$ column of \Cref{tab:lab_slopes}.
Each PMT shows a significant positive dependence of the gain on the magnetic field, with a tendency for PMTs with larger serial numbers to exhibit a stronger dependence, consistent with the trend suggested in the LST-1 data, as shown in \Cref{fig:SlopeVsSerial}.
These results support the interpretation that the gain dependence on the telescope pointing direction observed in the LST-1 data is caused by the geomagnetic field.

A linear fit to the laboratory data using all PMT samples yields a relative gain change of $0.066\pm 0.002\,\mathrm{mT}^{-1}$, which is smaller than that observed in the LST-1 data, $0.091\pm 0.001\,\mathrm{mT}^{-1}$.
The quoted uncertainties represent the statistical uncertainties of the linear fits only.
Although the difference is large compared with the statistical uncertainties, the statistical uncertainties do not reflect the substantial variation in the magnetic-field dependence among individual PMTs.
The LST-1 data indicate such a variation, as discussed in \Cref{sec:onsite_results}, and the PMT samples measured in the laboratory indeed show a wide range of magnetic-field dependence, from $0.032\pm 0.001\,\mathrm{mT}^{-1}$ to
$0.111\pm 0.001\,\mathrm{mT}^{-1}$.
Therefore, the limited number of PMT samples used in the laboratory measurements may introduce a bias relative to the average over the $\sim 1800$ PMTs installed in the LST-1 camera.

Including the systematic $B$ term in \Cref{eq:F-factor-method} moderately changes the measured magnetic-field dependence, but does not alter the overall conclusion.
Assuming a typical value of $B=0.03$, obtained from the LST-1 camera calibration, the fitted linear slope for the combined PMT samples decreases to $0.056\pm 0.004\,\mathrm{mT}^{-1}$.
Given that the laboratory measurements were performed with faster sampling than the LST-1 camera and without waveform digitization, using $B=0.03$ is expected to provide a conservative estimate of the effect of the $B$ term.

\begin{table}
    \centering
    \caption{Slopes of the linear fit to the dependence of each parameter on the magnetic field strength measured in the laboratory.
    Here, $G_F$ is the gain estimated with the excess-noise-factor method using the \MPE\ measurement, while $G$ and $F^2$ are the gain and excess noise factor obtained from the \SPE\ measurement, respectively.
    All gain and excess noise factor values are normalized to their values measured in the absence of the magnetic field generated by the coils.
    The uncertainties represent the statistical errors of the fits.}
    \label{tab:lab_slopes}
    \begin{tabular}{cccc}
        \hline
        PMT Sample& d$G_F$/d$B$ [mT$^{-1}$] & d$G$/d$B$ [mT$^{-1}$] & d$F^2$/d$B$ [mT$^{-1}$] \\
        \hline
        ZQ4058 & 0.032$\pm$0.001 & 0.09$\pm$0.06 & -0.12$\pm$0.04 \\
        ZQ5628 & 0.038$\pm$0.001 & 0.12$\pm$0.06 & -0.12$\pm$0.05 \\
        ZQ6295 & 0.111$\pm$0.001 & 0.13$\pm$0.03 & -0.05$\pm$0.03 \\
        ZQ7568 & 0.061$\pm$0.001 & 0.14$\pm$0.04 & -0.05$\pm$0.03 \\
        ZQ7980 & 0.091$\pm$0.001 & 0.13$\pm$0.05 & -0.06$\pm$0.04 \\
        Combined fit & 0.066$\pm$0.002 & 0.13$\pm$0.02 & -0.07$\pm$0.02 \\
        \hline
    \end{tabular}
\end{table}

\subsubsection{\SPE\ Measurement}

\begin{figure}
    \centering
    \includegraphics[width=0.9\textwidth]{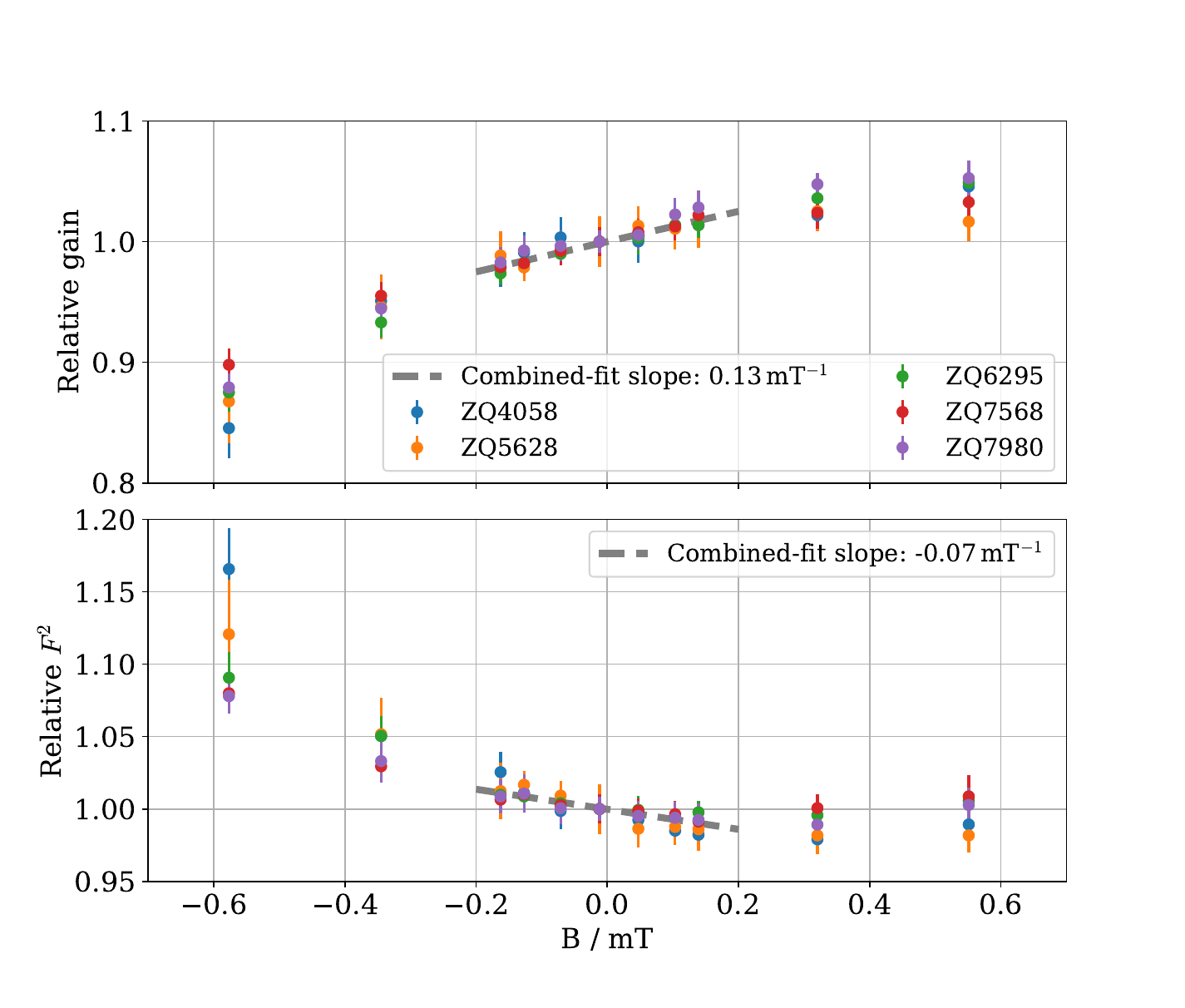}
    \caption{Dependence of the PMT gain, $G$, and the excess noise factor, $F^2$, on the magnetic field strength obtained from the \SPE\ measurement. Both quantities are normalized to the values obtained with no magnetic field applied by the coils. The gray dashed lines indicate linear fits to the data between $-0.2$\,mT and $0.2$\,mT using all PMT samples.}
    \label{fig:SPE_vs_B}
\end{figure}

\Cref{fig:SPE_vs_B} shows the dependence of the PMT gain and the excess noise factor on the magnetic field strength extracted from the \SPE\ charge spectrum analysis.
Since the results indicate a non-linear dependence in both quantities, a linear fit is applied to the data points in the range $|B|<0.2$\,mT to quantify their relative changes around $B=0$\,mT, as shown by the dashed lines in \Cref{fig:SPE_vs_B}.
When combining all the PMT samples, the PMT gain is suggested to be positively correlated with the magnetic field strength, with a slope of $0.13\pm0.02$\,mT$^{-1}$, whereas the excess noise factor shows a negative correlation with a slope of $-0.07\pm0.02$\,mT$^{-1}$.
The dependencies are also examined in each PMT sample, as summarized in \Cref{tab:lab_slopes}.

The PMT gain estimation by the excess-noise-factor method, $G_F$, using \Cref{eq:F-factor-method}, is dependent both on the actual PMT gain $G$ and the excess noise factor $F^2$ as
\begin{equation}
    G_{F}\propto GF^2.
\end{equation}
The \SPE\ measurement results thus expect a dependence of $G_F$ on the magnetic field strength as $\sim0.13{\rm \,mT^{-1}}-0.07{\rm \,mT^{-1}}=0.06{\rm \,mT^{-1}}$.
This is consistent with the $G_{F}$ dependence found in the \MPE\ measurement results ($0.066\pm0.002$\,mT$^{-1}$).
The consistency between the \SPE\ and \MPE\ results also holds for each of the PMT sample, though uncertainty in the \SPE\ results is relatively large.
Note that the consistency between the \MPE\ and \SPE\ measurements is observed even with different HV values applied, 1030--1080\,V for \MPE\ and 1500\,V for \SPE\ measurements.
This is in line with the expectation that the excess noise factor of the LST-1 PMTs is essentially independent of the HV value, as it is dominated by the electron multiplication at the first dynode, where the voltage is fixed.

While all the PMT samples give a similar dependence of their gain on the magnetic field, the dependence of the excess noise factor may vary among the different PMTs.
Especially, the samples ZQ4058 and ZQ5628 show steeper slopes, which may explain their smaller dependence of $G_F$ on the magnetic field observed in \Cref{fig:MPEGain_vs_B}.
The magnetic-field dependence of $F^2$ obtained in this study indicates that the average systematic variation of $F^2$ due to the geomagnetic field at the LST-1 site is at most $\sim$0.5\%, which is negligible compared to the precision required for the LST-1 calibration.

\section{Conclusions}

The impact of magnetic fields on the LST-1 PMTs was investigated using both on-site data and dedicated laboratory measurements.
The PMT gain estimated from the on-site data shows a dependence on the magnetic-field component along the axis perpendicular to the line connecting the first and second dynodes, as viewed from the photocathode.
This is consistent with an asymmetric modification of photoelectron trajectories with respect to the dynode geometry.
A similar dependence was reproduced in the laboratory using the \MPE\ measurements, supporting the interpretation that the gain variation observed in the LST-1 data is caused by the geomagnetic field.
Although the magnitude of the effect was not perfectly reproduced, the discrepancy may be explained by the limited number of PMT samples used in the laboratory measurements and differences in the magnetic field configuration between the LST-1 camera and the laboratory.
The \SPE\ measurements further revealed different dependencies of the PMT gain and the excess noise factor on the magnetic field, consistent with the \MPE\ results when considered together.
The variation of the excess noise factor due to the geomagnetic field was found to be $\sim0.5\%$, which is negligible for the LST-1 calibration.
The results provide a clearer understanding of the LST-1 observations in terms of the geomagnetic field.

\appendix

\section{Dependence of the PMT Gain on $B_y$ and $B_z$ during the On-site Test}  \label{sec:GainvsByBz}

\begin{figure}
    \centering
    \includegraphics[width=0.8\textwidth]{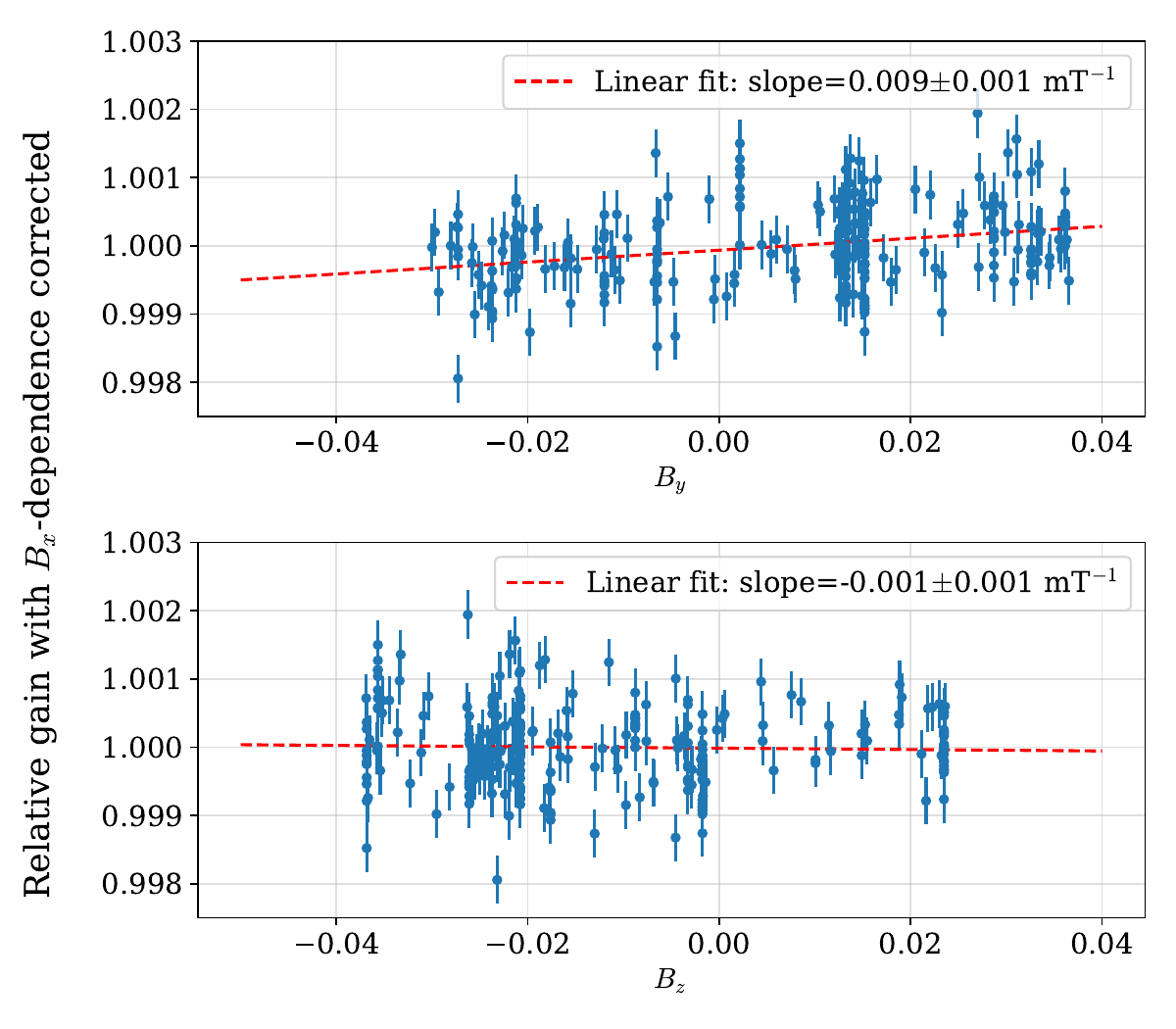}
    \caption{Dependence of the PMT gain on $B_y$ and $B_z$ obtained from the on-site data, after correcting for the $B_x$ dependence shown in \Cref{fig:GainvsBx_LST}. The red dashed lines indicate linear fits to the data. The error bars represent statistical error only.}
    \label{fig:GainVsByBz}
\end{figure}

To examine whether the PMT gain exhibits any additional dependence on the magnetic-field components other than $B_x$, the dependence of the PMT gain on $B_y$ and $B_z$ was also evaluated using the on-site data.
To remove the effect of $B_x$, the measured PMT gain was corrected by dividing it by the value expected from the fitted linear relation shown in \Cref{fig:GainvsBx_LST}.
\Cref{fig:GainVsByBz} shows the dependence of the corrected PMT gain on $B_y$ and $B_z$.
While the fitted slope is consistent with zero for $B_z$, a small dependence on $B_y$ was found, with a slope of $0.009\pm0.001$\,mT$^{-1}$.
This residual dependence may arise from incomplete removal of the $B_x$ dependence due to uncertainties in determining the local coordinate system of the PMTs.
However, the magnitude of this dependence is only about 10\% of the dependence on $B_x$, and does not affect the overall interpretation of the results.

\section{Dependence of the Geomagnetic-Field Effect on PMT Position in the Camera}  \label{sec:SlopecVsDist}

\begin{figure}
    \centering
    \includegraphics[width=0.8\textwidth]{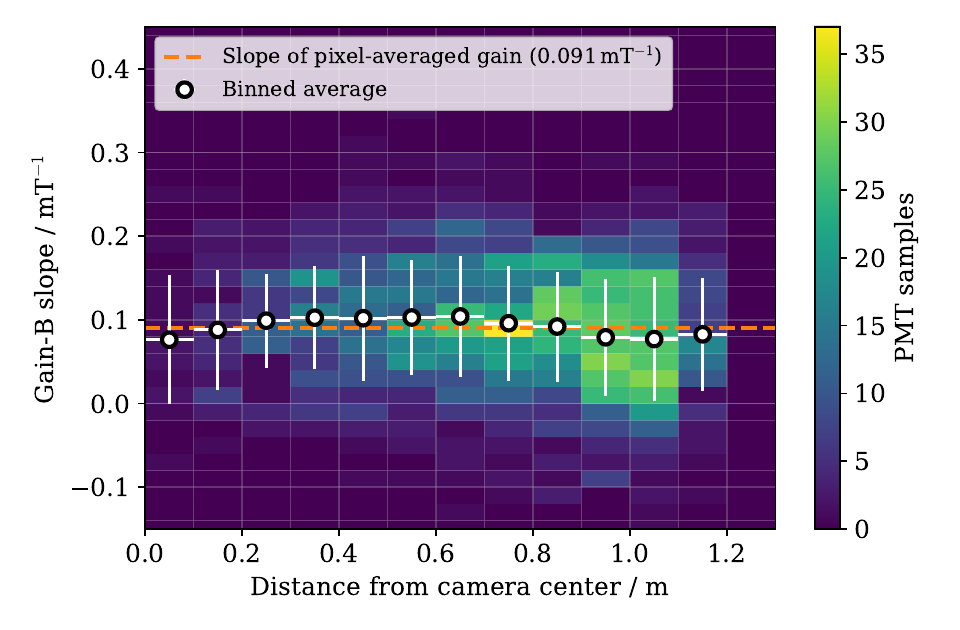}
    \caption{Distribution of the PMT samples analyzed in the on-site test as a function of the distance from the camera center and the fitted slope of the PMT gain as a function of the magnetic field. The white points represent the average slope in each distance bin, with error bars showing the standard deviation within each bin. The orange dashed line indicates the average dependence, corresponding to the red dashed line in~\Cref{fig:GainvsBx_LST}}
    \label{fig:SlopeVsDist}
\end{figure}

The magnetic field experienced by the PMTs during the on-site test could depend on the PMT position within the camera if the effective shielding provided by the mu-metal shields is not uniform across the camera or if there are additional magnetic-field sources inside the camera.
To examine these possibilities, the dependence of the observed geomagnetic-field effect on the PMT position was investigated.
\Cref{fig:SlopeVsDist} shows the fitted slope of the magnetic-field dependence as a function of the distance from the camera center.
No systematic trend with the PMT position is observed beyond the variation among individual PMTs.
Therefore, the observed variation in the geomagnetic-field effect among the PMTs cannot be explained by differences in their positions within the camera.

\section{Calibration of the Magnetic Field in the Laboratory Setup}  \label{sec:magcal}
\begin{figure}
    \centering
    \includegraphics[width=0.8\textwidth]{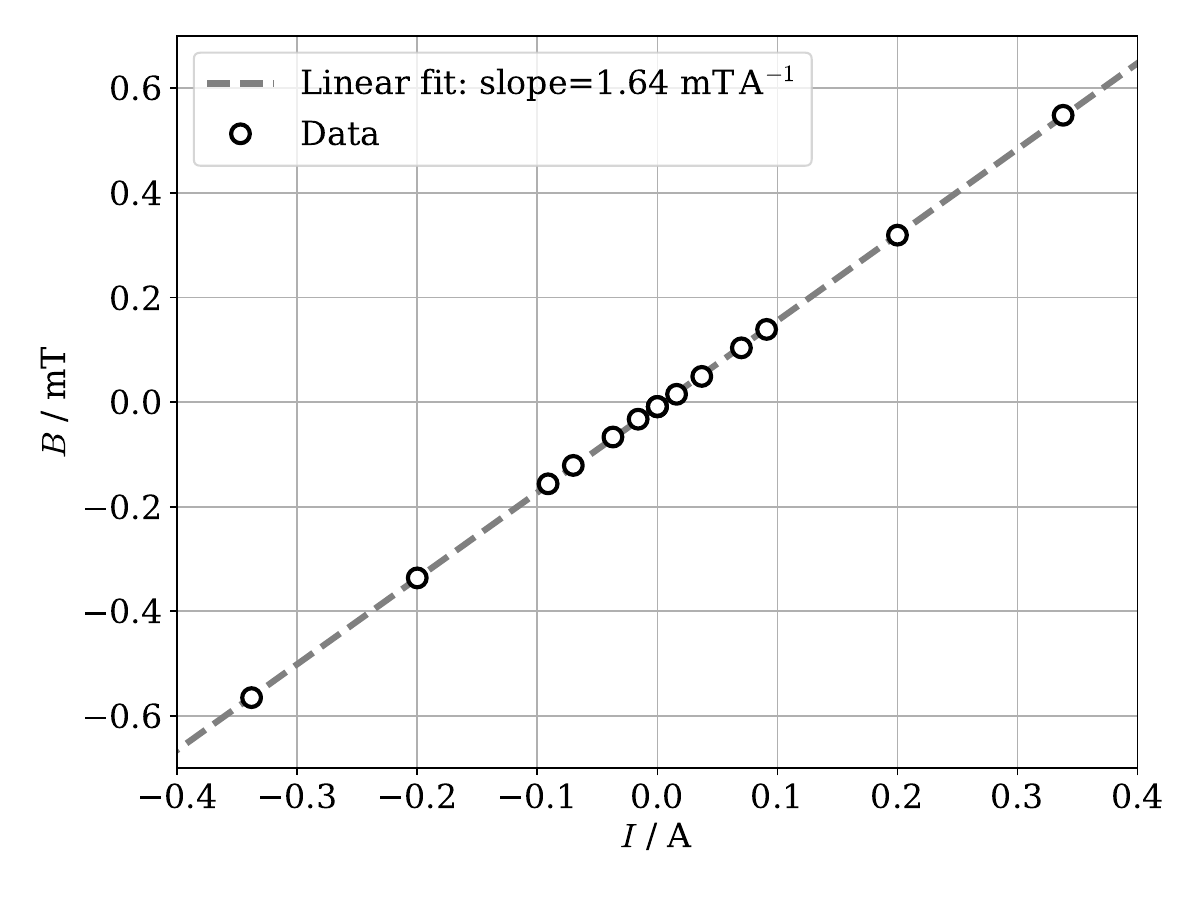}
    \caption{Relation between the current supplied to the Helmholtz coils and the magnetic field measured between the coils.}
    \label{fig:BvsI}
\end{figure}

In order to calibrate the magnetic field strength generated by the Helmholtz coils, the magnetic field was measured for several values of the current applied to the coils.
When a magnetic field $B$ is generated by the two coils carrying a current $I$, the field strength on the central axis can be calculated as
\begin{equation}
B = \left(\frac{4}{5}\right)^{\frac{3}{2}}\frac{\mu N I}{r},
\label{eq:helmholtz-coil}
\end{equation}
where $\mu$ is the permeability of air, and $N$ and $r$ are the winding number and the radius of the coils, respectively.
The magnetic field generated in our setup was measured using a Model~425 Gaussmeter from Lake Shore Cryotronics. 
The field was measured along the $x$-axis with the probe placed between the coils on the central axis.
\Cref{fig:BvsI} shows the relationship between the current supplied to the coils and the magnetic field measured at the coil center.
A clear linear relationship between $B$ and $I$ is observed.
Given the coil parameters of $N = 200$ and $r = 10.5$\,cm, the expected proportionality coefficient is 1.71\,mT\,A$^{-1}$.
The linear fit to the measured data yields a slope of 1.64\,mT\,A$^{-1}$, which is in good agreement with the expected value. 
The small discrepancy of $\sim4$\% may arise from deviations of the coils from the ideal Helmholtz configuration, such as distortions in the shape or uncertainties in the effective coil radius.
In addition, the probe position during the magnetic field measurement can introduce a systematic uncertainty.
The x-component of the geomagnetic field present in the setup was measured to be $B_0 = -0.01$\,mT. Since the geomagnetic field was not shielded during the measurements, the magnetic field strength is calculated including this contribution.

\section{Modeling of the \SPE\ Charge Spectrum}  \label{sec:spe_model}

This section describes the \SPE\ charge spectrum fit following~\cite{LST_Module}.
To fit the charge distribution obtained from the \SPE\ measurement in the laboratory, contributions from pedestal, \SPE, and two-p.e. events are considered.
The pedestal contribution, $m_0(q)$, is modeled by a Gaussian distribution, where $q$ is the PMT charge.
The pure \SPE\ contribution without pedestal noise, $m_1(q)$, is empirically modeled as a combination of a King function, $m^{\rm King}_1(q)$, and an adaptive power-law (APL) function, $m^{\rm APL}_1(q)$.
Each function is written as
\begin{eqnarray}
m^{\rm King}_1(q) &=& \frac{a}{\sigma^2}\left(1+\frac{1}{2\gamma}\left(\frac{q-\mu}{\sigma}\right)\right)^{-\gamma}\\
m^{\rm APL}_1(q) &=& 
\begin{cases}
b\left(1 - \frac{q}{s}\right)^{\alpha + (\beta-\alpha)q/s} &\text{if } q\leq s, \\
0 &\text{if } q>s,
\end{cases}
\label{eq:spe_funcs}
\end{eqnarray}
where $a$, $\mu$, $\sigma$, $\gamma$, $b$, $s$, $\alpha$, and $\beta$ are fitting parameters.
The pure two-p.e. contribution is obtained by convolving the \SPE\ model with itself,
$m_2(q) = m_1(q)\circledast m_1(q)$.
Taking into account pedestal noise in the \SPE\ and two-p.e. events, the practical contributions to the measured charge spectrum are
\begin{eqnarray}
M_1(q)=m_0(q)\circledast m_1(q), \quad
M_2(q)=m_0(q)\circledast m_2(q),
\end{eqnarray}
respectively.
The measured charge distribution can then be expressed as the sum of all contributions:
\begin{equation}
M(q) = N_0 \left( P^\lambda_0 m_0(q) + P^\lambda_1 M_1(q) + P^\lambda_2 M_2(q) \right),
\label{eq:spe_model_all}
\end{equation}
where $N_0$ is the total number of events, $\lambda$ is the average number of p.e. observed per event, and $P^\lambda_n$ is the Poisson probability of observing $n$ p.e.
The Poisson parameter $\lambda$ is included among the fitting parameters.

\section*{Acknowledgements}
This work was supported in part by the Ministry of Education, Culture, Sports, Science and Technology (MEXT), Japan, through funding allocated to the Institute for Cosmic Ray Research (ICRR), the University of Tokyo, as an International Joint Usage/Research Center, and by the Joint Research Program of ICRR. Additional support was provided by JSPS KAKENHI Grant Number JP23H05430.

\bibliography{mybibfile}

@article{HESS,
title = {Calibration of cameras of the H.E.S.S. detector},
journal = {Astroparticle Physics},
volume = {22},
number = {2},
pages = {109-125},
year = {2004},
issn = {0927-6505},
doi = {https://doi.org/10.1016/j.astropartphys.2004.06.006},
url = {https://www.sciencedirect.com/science/article/pii/S0927650504001227},
author = {F. Aharonian and others}
}

@ARTICLE{MAGIC,
       author = {{MAGIC Collaboration}},
        title = "{The major upgrade of the MAGIC telescopes, Part I: The hardware improvements and the commissioning of the system}",
      journal = {Astroparticle Physics},
         year = 2016,
        month = jan,
       volume = {72},
        pages = {61-75},
          doi = {10.1016/j.astropartphys.2015.04.004},
archivePrefix = {arXiv},
       eprint = {1409.6073},
 primaryClass = {astro-ph.IM},
       adsurl = {https://ui.adsabs.harvard.edu/abs/2016APh....72...61A}
}

@ARTICLE{VERITAS,
       author = {{Holder}, J. and others},
        title = "{The first VERITAS telescope}",
      journal = {Astroparticle Physics},
         year = 2006,
        month = jul,
       volume = {25},
       number = {6},
        pages = {391-401},
          doi = {10.1016/j.astropartphys.2006.04.002},
archivePrefix = {arXiv},
       eprint = {astro-ph/0604119},
 primaryClass = {astro-ph},
       adsurl = {https://ui.adsabs.harvard.edu/abs/2006APh....25..391H}
}

@article{LEONORA2013,
    title = {{Terrestrial magnetic field effects on large photomultipliers}},
    journal = {Nuclear Instruments and Methods in Physics Research Section A: Accelerators, Spectrometers, Detectors and Associated Equipment},
    volume = {725},
    pages = {148-150},
    year = {2013},
    note = {VLVnT 11, Erlangen, Germany, 12 - 14 October, 2011},
    issn = {0168-9002},
    doi = {https://doi.org/10.1016/j.nima.2012.12.056},
    url = {https://www.sciencedirect.com/science/article/pii/S0168900212015902},
    author = {E. Leonora},
}

@article{CALVO2010,
title = {{Characterization of large-area photomultipliers under low magnetic fields: Design and performance of the magnetic shielding for the Double Chooz neutrino experiment}},
journal = {Nuclear Instruments and Methods in Physics Research Section A: Accelerators, Spectrometers, Detectors and Associated Equipment},
volume = {621},
number = {1},
pages = {222-230},
year = {2010},
issn = {0168-9002},
doi = {https://doi.org/10.1016/j.nima.2010.06.009},
url = {https://www.sciencedirect.com/science/article/pii/S016890021001199X},
author = {E. Calvo and others},
}

@article{Inome2019,
  author={Inome, Yusuke and others},
  journal={IEEE Transactions on Nuclear Science}, 
  title={{A 100-ps Pulse Laser as a Calibration Source}}, 
  year={2019},
  volume={66},
  number={8},
  pages={1993-1997},
  doi={10.1109/TNS.2019.2928800}}

@article{Ritt2008,
  title={{Design and performance of the 6 GHz waveform digitizing chip DRS4}},
  author={Stefan Ritt},
  journal={2008 IEEE Nuclear Science Symposium Conference Record},
  year={2008},
  pages={1512-1515}
}

@article{sanuy2012,
  title={{Wideband (500 MHz) 16 bit dynamic range current mode PreAmplifier for the CTA cameras (PACTA)}},
  author={Sanuy, A and others},
  journal={Journal of Instrumentation},
  volume={7},
  number={01},
  pages={C01100},
  year={2012},
  publisher={IOP Publishing}
}

@article{f-factor-method,
title = {{A simple light detector gain measurement technique}},
journal = {Nuclear Instruments and Methods in Physics Research Section A: Accelerators, Spectrometers, Detectors and Associated Equipment},
volume = {315},
number = {1},
pages = {349-353},
year = {1992},
issn = {0168-9002},
doi = {https://doi.org/10.1016/0168-9002(92)90727-L},
url = {https://www.sciencedirect.com/science/article/pii/016890029290727L},
author = {B. Bencheikh and others}
}

@software{cassol_2025_15729582,
  author       = {Franca Cassol and Maximilian Linhoff},
  title        = {{lstcam\_calib: a python package to perform the
                   camera calibration of the Large-Sized Telescope
                   (LST) of the CTAO."
                  }},
  month        = jun,
  year         = 2025,
  publisher    = {Zenodo},
  version      = {v0.1.1},
  doi          = {10.5281/zenodo.15729582},
  url          = {https://doi.org/10.5281/zenodo.15729582},
}

@article{LST_Module,
title = {{Development and quality control of PMT modules for the large-sized telescopes of the Cherenkov Telescope Array Observatory}},
journal = {Nuclear Instruments and Methods in Physics Research Section A: Accelerators, Spectrometers, Detectors and Associated Equipment},
volume = {1073},
pages = {170229},
year = {2025},
issn = {0168-9002},
doi = {https://doi.org/10.1016/j.nima.2025.170229},
url = {https://www.sciencedirect.com/science/article/pii/S0168900225000300},
author = {T. Saito and others},
}

@article{LST1CamCal,
title = {{Camera calibration of the first Large-Sized Telescope of the Cherenkov Telescope Array Observatory}},
journal = {Astroparticle Physics},
volume = {175},
pages = {103189},
year = {2026},
issn = {0927-6505},
doi = {https://doi.org/10.1016/j.astropartphys.2025.103189},
url = {https://www.sciencedirect.com/science/article/pii/S0927650525001124},
author = {Franca Cassol and others},
}

@article{IGRF14,
	author = {Beggan, C. D. and others},
	date = {2026/06/29},
	doi = {10.1186/s40623-025-02360-0},
	id = {Beggan2026},
	isbn = {1880-5981},
	journal = {Earth, Planets and Space},
	number = {1},
	pages = {127},
	title = {International geomagnetic reference field: the fourteenth generation},
	url = {https://doi.org/10.1186/s40623-025-02360-0},
	volume = {78},
	year = {2026}
}

@article{CaliBox,
title = {{Test results of the optical calibration system for the Large Sized Telescope camera}},
journal = {Astroparticle Physics},
volume = {167},
pages = {103079},
year = {2025},
issn = {0927-6505},
doi = {https://doi.org/10.1016/j.astropartphys.2025.103079},
url = {https://www.sciencedirect.com/science/article/pii/S0927650525000027},
author = {M. Iori and others},
}

@article{LSTPMT2016,
title = {{Evaluation of Photo Multiplier Tube candidates for the Cherenkov Telescope Array}},
journal = {Nuclear Instruments and Methods in Physics Research Section A: Accelerators, Spectrometers, Detectors and Associated Equipment},
volume = {824},
pages = {640-641},
year = {2016},
note = {Frontier Detectors for Frontier Physics: Proceedings of the 13th Pisa Meeting on Advanced Detectors},
issn = {0168-9002},
doi = {https://doi.org/10.1016/j.nima.2015.08.030},
url = {https://www.sciencedirect.com/science/article/pii/S016890021500964X},
author = {R. Mirzoyan and others},
}

@article{LSTPMT2017,
title = {{Evaluation of novel PMTs of worldwide best parameters for the CTA project}},
journal = {Nuclear Instruments and Methods in Physics Research Section A: Accelerators, Spectrometers, Detectors and Associated Equipment},
volume = {845},
pages = {603-606},
year = {2017},
note = {Proceedings of the Vienna Conference on Instrumentation 2016},
issn = {0168-9002},
doi = {https://doi.org/10.1016/j.nima.2016.06.080},
url = {https://www.sciencedirect.com/science/article/pii/S0168900216306416},
author = {R. Mirzoyan and others},
}

@article{Lightguide2017,
doi = {10.1088/1748-0221/12/12/P12008},
url = {https://doi.org/10.1088/1748-0221/12/12/P12008},
year = {2017},
month = {dec},
publisher = {},
volume = {12},
number = {12},
pages = {P12008},
author = {Okumura, A. and others},
title = {{Prototyping hexagonal light concentrators using high-reflectance specular films for the Large-Sized Telescopes of the Cherenkov Telescope Array}},
journal = {Journal of Instrumentation},
}

@misc{LST_PMT_Inome,
author="Yusuke Inome and others",
title="Development of the camera for the Large Size Telescopes of the Cherenkov Telescope Array",
journal="Proc. SPIE",
year="2014",
volume="9151",
pages="915142",
URL="https://cir.nii.ac.jp/crid/2120589441485213696"
}

\end{document}